*Article*

# Percolation Diffusion into Self-Assembled Mesoporous Silica Microfibres

**John Canning,**\*,1,4**, George Huyang**[1]**, Miles Ma**[1]**, Alison Beavis**[2]**, David Bishop**[2]**, Kevin Cook**[1]**, Andrew McDonagh**[2]**, Donqi Shi**[3]**, Gang-Ding Peng**[3] **and Maxwell J. Crossley**[4]

1   *i*nterdisciplinary Photonics Laboratories, School of Chemistry, The University of Sydney, NSW 2006 Australia
2   School of Chemistry & Forensic Science / Institute for Nanoscale Technology, University of Technology Sydney, NSW 2007 Australia
3   Photonics & Optical Communications, School of Electrical Engineering and Telecommunications, The University of New South Wales, NSW 2052, Australia
4   School of Chemistry, The University of Sydney, NSW 2006 Australia

\*   Author to whom correspondence should be addressed; E-Mail: john.canning@sydney.edu.au; Tel.: +61-2-93511934.

**Abstract:** Percolation diffusion into long (11.5 cm) self-assembled, ordered mesoporous microfibres is studied using optical transmission and laser ablation inductive coupled mass spectrometry (LA-ICP-MS). Optical transmission based diffusion studies reveal rapid penetration (< 5 s, $D > 80$ μm$^2$.s$^{-1}$) of Rhodamine B with very little percolation of larger molecules such as zinc tetraphenylporphyrin (ZnTPP) observed under similar loading conditions. The failure of ZnTPP to enter the microfibre was confirmed, in higher resolution, using LA-ICP-MS. In the latter case, LA-ICP-MS was used to determine the diffusion of zinc acetate dihydrate, $D \sim 3 \times 10^{-4}$ nm$^2$.s$^{-1}$. The large differences between the molecules are accounted for by proposing ordered solvent and structure assisted accelerated diffusion of the Rhodamine B based on its hydrophilicity relative to the zinc compounds. The broader implications and applications for filtration, molecular sieves and a range of devices and uses are described.

## 1. Introduction

The recent cold fabrication of optical microfibre waveguides from silica nanoparticles using a combination of self-assembly and controlled fracturing[1-3] offers a route to a range of applications exploiting new components and waveguides, single photon sources for quantum communications, potential interconnect applications, specialty sensors, molecular sieves and filters and so on. The approach is generic and was recently used to fabricate dye doped poly(methylmethacrylate) (PMMA) nanoparticles[4]. The regime being examined is truly nanoscale and differs from polystyrene work on a sub-micron scale[5-8] often used to self-assemble, for example, artificial opals with periodic structures on a scale commensurate with Bragg scattering in the visible. Convective, microfluidic flow within a pinned drop during evaporation along with attractive intermolecular forces first leads to ordered, "spontaneous" self-assembly or packing of the nanoparticles into a 2-D film on an amorphous substrate. Self-organisation occurs despite being on a hydrophilic, amorphous borosilicate surface with inhomogeneous chemical and surface topology, suggesting that the role of template driven self-assembly is not significant. Rather, the driver is a tendency towards the densest possible state, or lowest free energy[9] configuration, which for classical hard spheres are hcp and fcc lattices[10]. More generally, this driver was observed to apply in the self-assembly of long chain molecular structures into monolayer films, such as 5,10,15,20-tetralkylporphyrins, which had been assumed to be template driven by the pyrolytic graphite substrate[11]. An analysis of X-ray diffraction of the apparently random precipitation of the molecule in solution turned out to have an identical, layered packing structure in three dimensions[12]. Thus there is experimental evidence to indicate that the nanoparticle system appears to have an analogy with molecular topology driven self-assembly above other environmental and chemical contributions. In either case, room temperature fabrication is a significant advantage compared to other means of fabricating microfibres, such as high temperature (> 1700 - 1900 °C) drawing of optical fibres[13], which are themselves drawn from fused silica preforms, or spun at below 1000 °C directly from sol gel glass[14]. The direct interaction of near universal attractive forces relies on proximity and can be utilized with efficiency at room temperature to create material composite systems not previously possible - the integration of nitrogen vacancy (NV) containing nanodiamonds directly into silica via self-assembly and demonstrating single photon emission within glass was possible[1]. Otherwise, these centres anneal out at 700 °C preventing the use of conventional methods. The reported work demonstrates a significant improvement over previous room temperature, surfactant based self-assembly of silica particles at water-air or water-liquid interfaces, including that used to fabricate silica fibres up to 5 cm long where hydrogen bonding, a similar dispersive force, is thought to be the key glue involved[15].

3-D slab microfibres result from fracturing of the self-assembled structure when the drop begins to finally recede during evaporation – the radial component of drying is not aligned with packing structure and stresses arise. For large round drops, this involves bifurcation, sometimes multiple times, in fracture before the waveguide structure becomes uniform, defined by an effective taper aspect ratio[1], TAR = 1. That this type of fracturing occurs is early evidence of this packing towards the densest possible state (hexagonal in the plane of the substrate and hcp/fcc otherwise). By controlling microfluidic flow within a drop it is possible to obtain unprecedented uniformity over the entire lengths ($L > 11$ cm), and aspect ratios in excess of $L/w > 15\,000$. Laser based processing of surfaces to pattern the degree of hydrophilicity or hydrophobicity demonstrated high finesse control over such microfluidic flow[2]. Scanning electron microscopy (SEM) and atomic force microscopy (AFM) showing height variations within a single nanoparticle diameter, together with agreement between simple modeling and gas adsorption measurements of pore diameters, supports the presence a highly ordered structure[3], indicative of either hcp or fcc packing. The idealised structures are therefore expected to be made up of interconnected nanochambers of at least two sizes (tetrahedral and octahedral lattice sites), illustrated in Figure 1, and percolation corresponds with diffusion through these sites. From standard packing theory of crystals, the ideal pore volume in such lattices is typically 26 % and both the calculated, (2.2 – 6.2) nm, and measured pore sizes using gas adsorption, ~ (2 – 6) nm, are consistent with that[3]. These figures are, for example, comparable with Corning VYCOR, a porous glass used in chromatography[16]. This particular mesoporous range is interesting as it is larger than microporous (< 2 nm), often used for single or few layer molecular sieves[17], but sufficiently small enough that over the thickness of the microfibres selective diffusivity through twisting and interlinked nanochambers (Figure 1) should occur with potentially very high fidelity nonetheless, particularly for those molecules that are not spherical. These kinds of mesoporous structures have many potential applications, and in particular there has been significant emphasis on environmental sensing of highly toxic materials such as heavy metals[18]. Multiple microfibre filaments can also be bundled together for numerous macro applications as well as used individually as slab waveguides for optical propagation and photonic applications (both linear and nonlinear). Given the potential of integrating photonics into silica mesoporous structures, and therefore opening up new diagnostic approaches such as optical chromatography and other diffusive applications, as well as filtering for sensing, the study of percolation diffusion is an important aspect reported here.

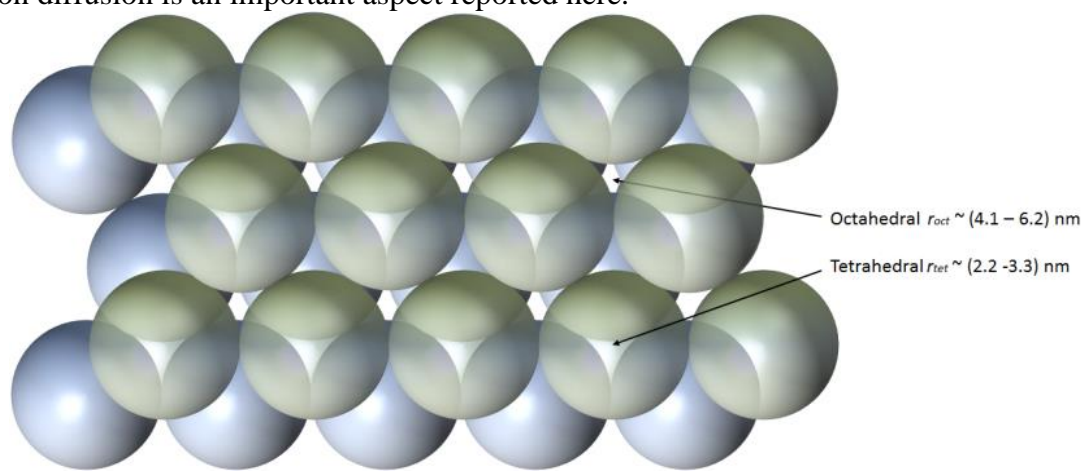

**Figure 1.** An illustration of hcp packing and the two types of interstitial regions. As an approximation a round sphere that can fit into the pores is often assumed to determine the size of a molecule that can percolate or diffuse into the structure. This does not, however, take into account irregular shapes and quantities of larger molecules, more typical of real situations. The site size distribution shown reflects that calculated from the bulk of the size distribution of the nanoparticles used in this work, measured by dynamic light scattering (DLS).

In this work, two methods are used to investigate these processes: (1) direct optical transmission measurements through the optical microfibres and (2) laser ablation inductive coupled mass spectrometry (LA-ICP-MS)[19-21]. The percolation, or diffusion, of three molecules is studied: Rhodamine B, zinc tetraphenylporhyrin (ZnTPP) and hydrated zinc acetate, summarised in Figure 2. Their longest dimension for each is 1.77, 1.74 and 0.82 nm respectively. ZnTPP has a square flat shape and is an uncharged molecule. Rhodamine B is cationic and contains a carboxylic group capable of binding to silica, whilst the chloride can be displaced by negatively charged interfacial water. Zinc acetate is significantly smaller than the other compounds examined here and is hydrated. The solvent was de-ionized water in all cases.

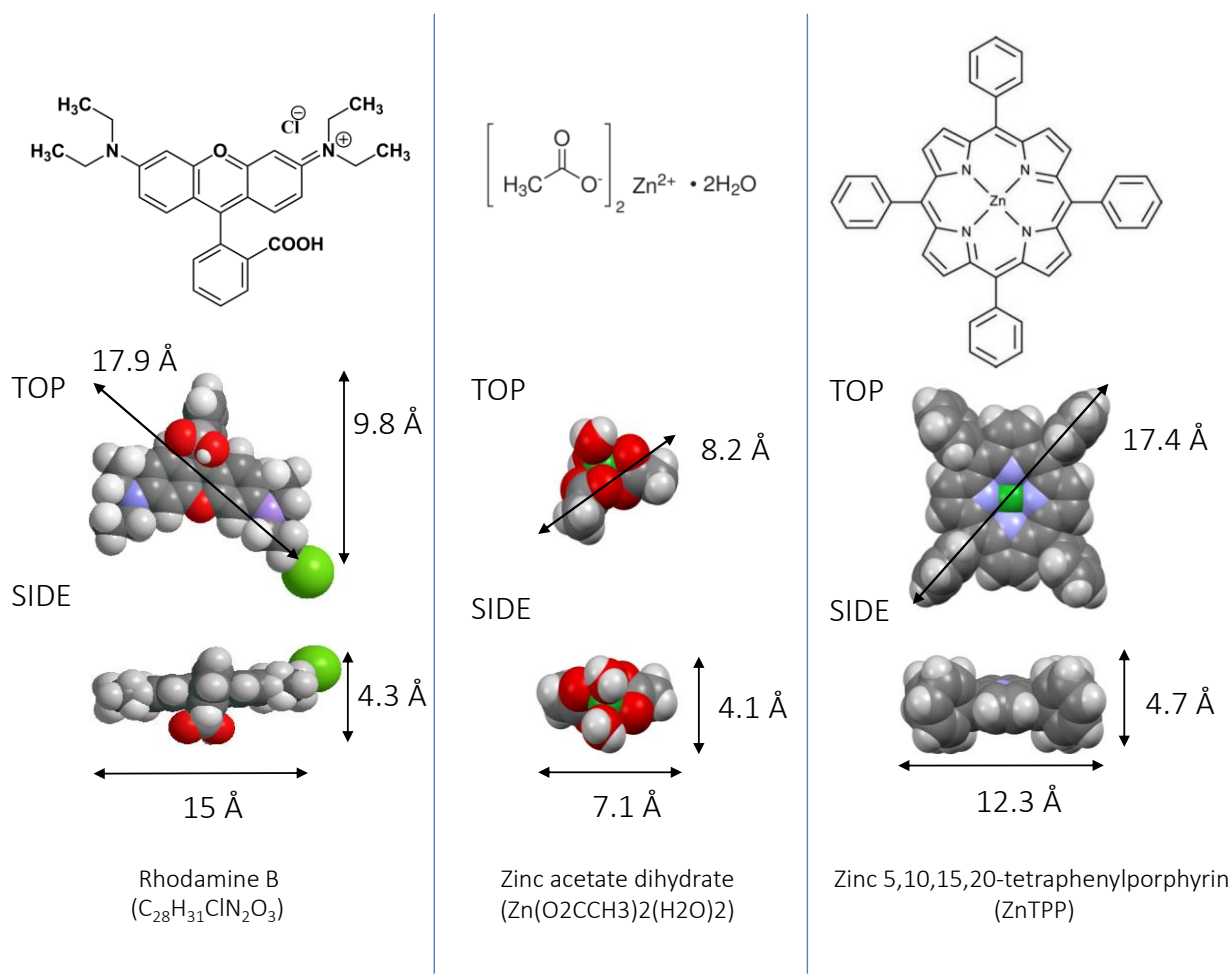

**Figure 2**. The formulas, schematic and space filling structures as well as dimensions of each of the three molecules used in this work. The Cl⁻ anion of Rhodamine B is free in solution and may be displaced by the negatively charged water at a silica-water interface.

## 2. Results and Analysis

### 2.1 Fabricating self-assembled, long silica microfibres

11.5 cm long silica microfibres with cross-sections of $w \times h = 40$ μm $\times$ 10 μm for this work were fabricated by gravity assisted directional evaporative self-assembly ("GADESA") where the substrate upon which the process is undertaken is tilted slightly. This generates a preferential convective flow in one axis within the larger drops helping to generate more uniform and aligned microfibres. A photograph of the microfibres is shown in Figure 3 (a). The overall structure of these microfibres is similar to that reported previously. For reference, Figure 3 (b) to (d) shows previous SEM analysis where the wires appear to have a top layer, ~ (250 – 500) nm in thickness, that appears distinct to that below suggesting different sized particles or a less dense finite number of layers on top of a much more dense inner structure. The penetration of external matter, however, cannot be excluded. The overall fcc and hcp packing at the surface is verified by AFM on the current fibres, shown in Figure 3 (e) – thus there is no observable difference in outcome using GADESA, indicating laminar flow within the drops. The question about the validity of extending the uniformity of structure at the surface to that of the volume and indeed everywhere else along the microfibre given that both SEM and AFM are highly localized techniques was qualitatively addressed in other work using gas adsorption studies[3] where pore size distribution of (2-6) nm agrees with calculations. These allowed a better assessment overall of the quality of structure through indirect comparison with expected pore volume and sizes for such lattices. Within error there was excellent agreement between measurement and calculation suggesting a homogenous and uniform inner structure was present.

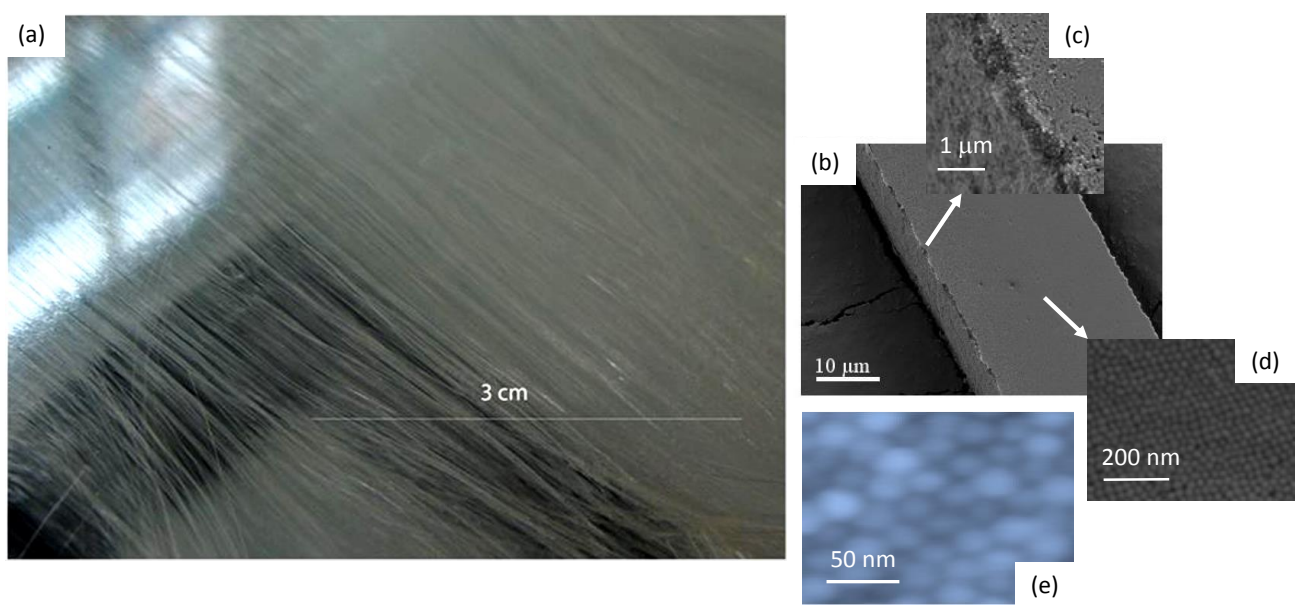

**Figure 3.** Surface characterisation of a self-assembled silica microwire: (a) optical micrograph of long wires produced using gravity assisted deposition by evaporative self-assembly; (b) – (c) SEM local images of the wire surface reproduced from earlier work (Naqshbandi *et al.* [1]) to illustrate the presence of a finite thickness layer, ~ (250 – 500) nm, (a) & (b) on top of an inner core. An examination of the surface layer (c,d) reveals what appears to be a mix of fcc and hcp packing in places; (e) shows an AFM analysis of the >11 cm wires used in this work with similar hcp packing to that of previous work [3].

*2.2 Diffusion measured by optical transmission*

The self-assembled wires are of sufficient dimension and high index contrast with the surrounding air-cladding to act as highly multimode slab (rectangular) waveguides. When light is launched into several angular rays[22,23], after propagating a few centimeters there is sufficient cross-coupling between all modes, aided by interface scattering, to ensure even coupling into all modes of the waveguide at the point at which optical interrogation is undertaken. A schematic of the experimental setup is shown in Figure 4. Essentially the transmission of light is monitored directly from the microfibre output to reduce multimode interference effects[24]. Both the Rhodamine B and the ZnTPP absorb at this wavelength so it is possible to determine their presence and percolation diffusion within the waveguide.

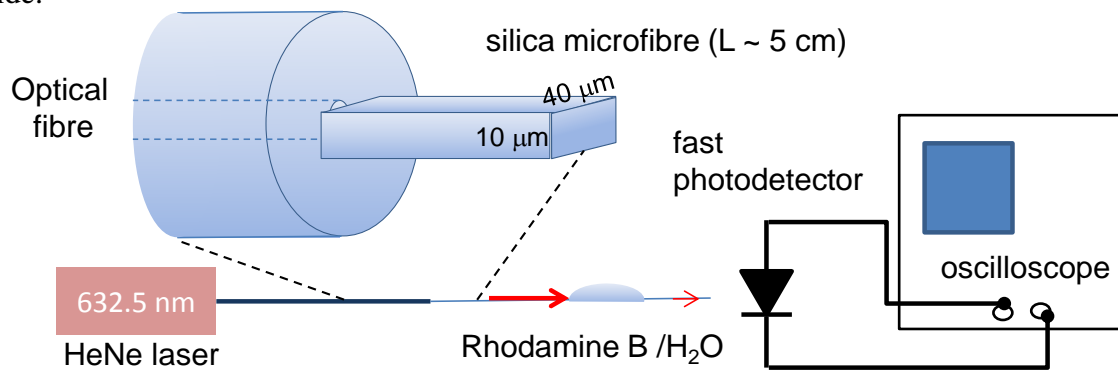

**Figure 4.** Schematic of the optical transmission measurement through the microfibre with a drop of Rhodamine B containing water. The coupling area between standard fibre and slab microfiber is zoomed in for clarity.

Figure 5 shows the transmission experiment results. Two notable features that were not possible to completely remove are the initial rise in transmission which occurs with impact of the drop on the wire and subsequent oscillations after very rapid penetration. The latter effects may be attributed to waveguide distortions arising from the addition of Rhodamine B and solvent, water, into the structure. Without considering Rhodamine B, as water diffuses into the structure and assuming ~ 26% to be air initially, the local real dielectric constant is given by superposition to be $\varepsilon_{(SiO2/x)} = \eta_{SiO2}.\varepsilon_{SiO2} + \eta_x.\varepsilon_x$, where $\eta$ is the fraction of component making up the structure. This will change the refractive index from $n_{(SiO2/Air)}$ ~ 1.35 to $n_{(SiO2/water)}$ ~ 1.42. This then suggests that there is no impediment to water diffusion and the Rhodamine B probably diffuses as fast as the solvent. As well as the dye absorbing light (imaginary index) it raises the local real part of the index generating an asymmetric waveguide and therefore altering some of its wave guiding properties. Other complications can exist including local heating due to absorption and other nonlinear effects, but essentially the region at which percolation occurs is no longer matched to regions outside this (until 5 s have passed) and so additional dynamic interference effects can lead to increased scattering of light and oscillations seemingly appear with time.

Nonetheless, the initial attenuation is approximately linear. The overall diffusion time into the whole waveguide, across the 10 μm thickness, (saturated after 20 to 30 s) is assumed to be $t < 5$ s. However, initial rapid intake before the onset of oscillations points to the diffusion time being significantly shorter ($t < 180$ ms). From the inset of Figure 5 even this relatively fast diffusion appears complex and the initial stage is exponential. Rhodamine B is a reasonably sized molecule, ~ 1.79 nm at its longest length[25]. Based on the measured times and assuming an exponential diffusion from one side across to the other we can estimate a diffusion coefficient $D \sim 80$ µm$^2$.s$^{-1}$ for a 5 s time, or $D > 2000$ µm$^2$.s$^{-1}$ for the faster time of $t < 180$ ms. On the other hand, if the diffusion is assumed to be from all around the waveguide, then this diffusion rate is probably up to four times slower. By comparison, the diffusion coefficient of Rhodamine 6G which is a little larger, for example, was measured in mesoporous structures[26] that had unimpeded channels with smaller diameters $\phi \sim 3.6$ nm to be $D \sim 3$ nm$^2$.s$^{-1}$, orders of magnitude slower.

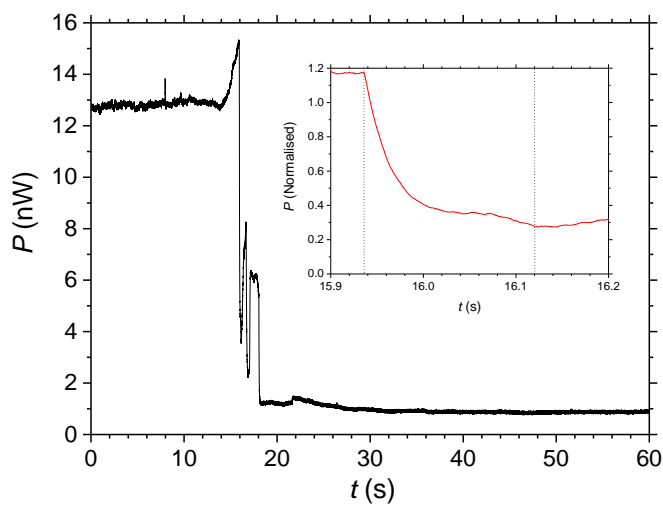

**Figure 5.** Optical transmission of 632 nm through a self-assembled wire as a function of time during percolation of Rhodamine B using the setup illustrated in Figure 4. Inset shows a close-up of the initial attenuation the signal has been normalized to the initial signal level measured.

In contrast to Rhodamine B, no diffusion of ZnTPP is observed at all. One explanation might be that its flat shape clogs and self-obstructs it passage around the bends between interstices; however, given the similar dimensions at the longest points of these species along with the speed with which the Rhodamine does penetrate, whilst probably a contributor this alone is not entirely satisfactory. The inability to percolate into the system suggests the porphyrin is effectively filtered out of the solvent. To study this in more detail and to rule out any role of complex waveguide responses to the material introduction (which need consideration if implementing all optical waveguide interrogation such as optical chromatography, optophoresis, molecular sieves with optical diagnostics and so on), a more direct approach of measuring the percolated material is employed.

*2.3 Measuring diffusion through laser ablation of self-assembled wires*

Laser ablation inductive coupled mass spectrometry (LA-ICP-MS) has previously been used to successfully track the presence of $Zn^{2+}$ and other metal ions in both polymer[27] and silica[28] glasses. LA-ICP-MS uses a UV laser to ablate surfaces to generate aerosols, which are collected using a carrier gas for elemental analysis by standard mass spectrometry. It is therefore also readily amenable to detecting trace metal ions such as $Zn^{2+}$ in a variety of biological[29] and material systems. In this experiment, zinc tetraphenylporphyrin ([ZnTPP] = *c* =100 mM) was allowed to sit on a wire (40 x 10 µm) from one side for 10 mins before LA-ICP-MS was undertaken. This time is well beyond the 5 s required to percolate Rhodamine B.

The ZnTPP is a nearly flat structured molecule with Zn(II) sitting in the centre of the conjugated ring structure (Figure 1). It is approximately 1.74 nm across its longest diagonal, a little smaller than the Rhodamine B at its longest point. This is a size approaching the smaller sized tetrahedral pores of the idealised silica microfibre structure ($\phi \sim$ 2-3 nm for 20 - 30 nm sized nanoparticles). Whilst a single molecule in a low concentration solution should be able to navigate its way into the structure through the larger pores, for clusters of such molecules trying to enter, we speculate that the large asymmetry and flat structure could lead to aggregation and clustering (clogging) within pores early on impeding diffusion or percolation. The failure to observe any attenuation with optical transmission experiments suggests almost no diffusion. It would appear to be drastically different to the similar sized Rhodamine B – a speculation might be for the Rhodamine B to be diffusing so readily, the orientation is clearly along it length when entering the pores. For comparison with this method, the smaller and more spherical-like hydrated zinc acetate ([zinc acetate] = 100 mM), was also used under identical conditions – in this case, the dopant diffuses directly into the wire and a greater flow is expected given it is nearly half the size of the ZnTPP.

The raw data as a function of time for both the zinc porphyrin and zinc acetate in water is shown in Figure 6 (a). It is evident that there is little penetration of the ZnTPP into the wires, much less than for the hydrated zinc ions, consistent with that observed in the optical transmission measurements. Compared to the Rhodamine B, both species diffuse significantly more slowly with the zinc acetate penetrating < 2 µm into the wire after 10 mins of exposure. Molecules are simply unable to pass by each other, something characteristic of normal mode diffusion within nanoporous structures[30], suggesting that in this case the effective mean free path is less than the pore dimensions, a situation approximating classical diffusion if somewhat impeded. Between the two zinc species, it may be concluded that whilst the porphyrin dimensions are smaller than the pores, they are sufficiently large when combined with their high shape asymmetry that clustering or aggregation prevents any significant penetration into the wire. By contrast, the hydrated Zn(II) ions are observed to enter the wire for the time of exposure used – this may be explained predominantly by the difference in size and shape. Nonetheless, its diffusion is far smaller than that of Rhodamine B suggesting that, despite hydration, there is little solvent involvement.

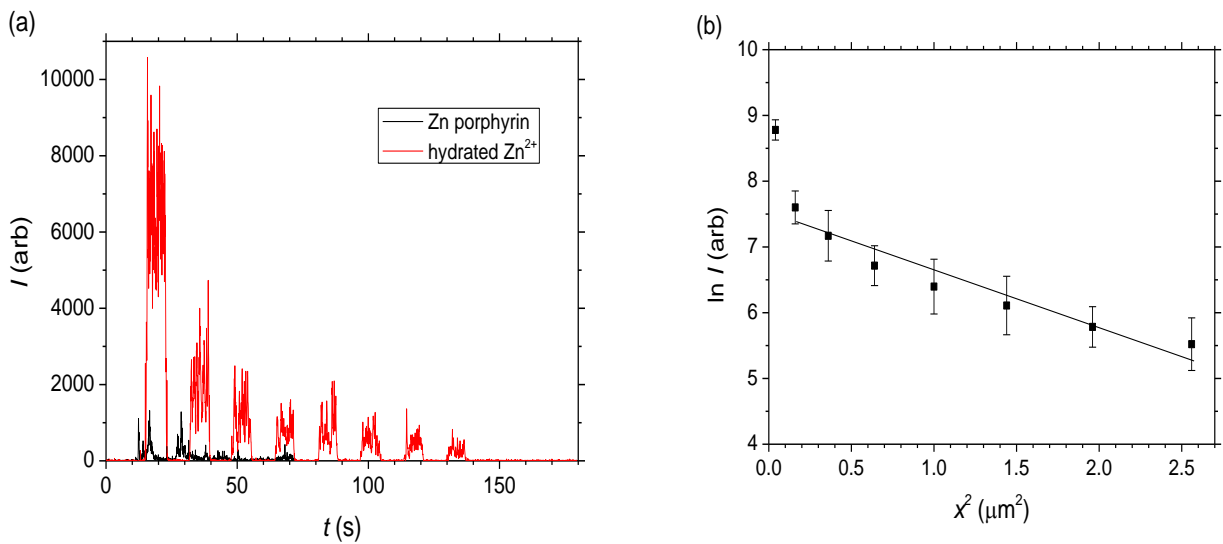

**Figure 6.** LA-ICP-MS of self-assembled wires doped with ZnTPP & zinc acetate solutions ($c_0$ = 100 mM). (a) Signal intensity as a function of each ablation over time for both molecules and (b) diffusion analysis (natural log of signal intensity vs area) of the hydrated $Zn^{2+}$ from the zinc acetate solution.

To calibrate the penetration depth, tilted SEM imaging of the ablation regions of a standard silica self-assembled wire was used as a reference. For 20 passes of the laser under identical conditions, the laser etching depth was determined to be ~ 4 μm. The final data for the zinc acetate solution is shown in Figure 6 (b). The decay is well fitted by a single exponential, treated as molecular or Fickian diffusion (the impeded flow assumed to arise from a confinement-enhanced Brownian motion which reduces the mean free path). Given the slab geometry of the wire, where the width is larger than the height ($w$ = 40 μm > $h$ = 10 μm), a simple one-dimensional diffusion is assumed as a first approximation[31]:

$$C = \frac{c_0}{A\sqrt{\pi D t}} e^{-\left(\frac{x^2}{4Dt}\right)} \qquad (1)$$

where $C$ is the number of Zn(II) ions and is proportional to the signal intensity $I$, $A$ is the area of the drop on the waveguide ($A$ ~ 0.5 cm x 40 μm), $D$ is the diffusion coefficient, $t$ is the time ($t$ = 10 min), and $x$ is the diffusion length into the wire. Therefore, the following expression can be plotted:

$$\ln(I) = \ln\left(\frac{I_0}{A\sqrt{\pi D t}}\right) - \frac{x^2}{4Dt} \qquad (2)$$

From the slope of this expression, an estimated mass spectrometer intensity based diffusion coefficient is $D$ ~ 4.8 x $10^{-4}$ $nm^2 s^{-1}$. Since $I \propto C = c.N_A$ where $N_A$ is Avogadro's number, the diffusion coefficient for the hydrated $Zn^{2+}$ can also be extracted from the slope of ln$C$ vs $x^2$ by converting to $C$ noting $I_0$ obtained from extrapolating to $x^2$ = 0 in Figure 6(b) is assumed to equal 0.1 M x 6.022 x$10^{23}$. From this, the effective molecular diffusion coefficient is estimated to be a little lower: $D$ ~ 3 x $10^{-4}$ $nm^2 s^{-1}$. Overall it is clear that even for the hydrated $Zn^{2+}$ which are small compared to the pore radius, liquid diffusion or percolation is very low compared to the Rhodamine B (and indeed compared to previous work with Rhodamine 6G in smaller channels[26]).

The fit in Fig 6(b) shows some nonlinearity which may also arise from a combination of diffusion from around the wire at each side and the presence of two distributions of pore sizes leading to a much more dispersive diffusion profile (better fitted by a stretched exponential or two exponentials if the pore sizes arise from identical spheres). Notwithstanding the 1-D approximation and the uncertainty arising from factors such as non-uniform ablation across the waveguide, which can be reduced by repeated measurements and more stable beam energies, the method of LA-ICP-MS is demonstrably an effective tool for characterising the penetration of species into self-assembled structures. For zinc acetate, the slow diffusion is described by Fickian diffusion and remains consistent with a classical random walk description of diffusion despite some relative confinement within the lattice structure.

## 3. Discussion

**Table 1**. Summary of effective diffusion coefficients for the three molecules used in this work.

| **Molecule** | Rhodamine B | Zinc acetate dihydrate | ZnTPP |
|---|---|---|---|
| **$D$ (m$^2$s$^{-1}$)** | $\sim (80 - 2000) \times 10^{-12}$ m$^2$s$^{-1}$ | $\sim (3 - 4.8) \times 10^{-22}$ m$^2$s$^{-1}$ | $< 10^{-22}$ m$^2$s$^{-1}$ |

Table 1 summarises the estimated effective diffusion rates, $D$, for each species. The observed diffusion of Rhodamine B clearly stands out relative to both $Zn^{2+}$ based compounds. Although size and shape of the porphyrin can likely explain why it is much slower than the hydrated zinc acetate, the difference with Rhodamine B is not so clear and is suggestive of the key role the solvent plays in diffusion. This is further complicated when comparing a commonly used definition for ballistic diffusion[32] where $\alpha = 2$ in the general diffusion relation, $<(x(t) - x(0))^2> \sim Dt^\alpha$. For Rhodamine B, from the transmission measurements $\alpha \gg 2$ which is suggestive of strong non-Brownian motion[33]. So a simple physical percolation of the molecules into passive, or equilibrated, channels within the wires based on Brownian motion within confined nanopores appears insufficient to account for the differences. These observations raise important issues with regards to the nature of the diffusive process, pertinent if only for the fact that dyes such as Rhodamine B that do penetrate into porous materials are attractive fluorescent markers commonly used for diffusion studies, including as target test drugs in studying potential drug release scaffolds[34]. Within these systems, where pore sizes are much larger than the mesoporous region reported here, a reported assumption that the Rhodamine B interactions are entirely confined to classical rheological interplay and diffusion seems valid. Interestingly, for scaffolds with equivalent pore dimensions well into the micron scale, the out-diffusion of the test "drug" Rhodamine B is more than 11 days[34], attributed to adsorption of the dye both in and out of the scaffold itself. Is adsorption assisting the super diffusive penetration of Rhodamine B into the microfibres?

It is clear for rapid percolation of the Rhodamine, its orientation must be almost aligned lengthwise such that the molecule's smallest dimensions are facing the direction of flow. This points to a preferred orientation and therefore some interaction with the environment to ensure it stays this way. The difference in charge between the molecules would suggest a difference in hydrophilicity with silica and the solvent, water: both zinc molecules have a positive charge held partially in place with a localised negative charge either on nitrogen (ZnTPP) or a carboxylic group (zinc acetate dihydrate). Therefore, there are no substantial solvent interactions and it would seem that diffusion can be described as classical, determined by rheological processes and limited by random clustering. In contrast, the Rhodamine B has a chloride anion that can be easily released in water leaving behind a positively charged nitrogen group that is attracted to the negatively charged interfacial water. This interaction sees Rhodamine B dragged along by the water and its orientation potentially determined by the orthogonal aromatic group with the carboxylic end group, which may also interact with silanated species on the silica surface – Brownian (thermal) motion within the water has to be effectively damped for this to occur suggesting the interfacial water is structurally self-organised and oriented itself. Ordering of interfacial water is often present in biological systems[35,36]. This interfacial water may even be made up of ice-like layers if deprotonation has occurred, with positive charges bound to negatively charged layers. In such a case, rapid solvent penetration may occur in part due to slip-assisted, or lubricated, diffusion – this is an example of a structure-assisted diffusion. It can be concluded that the Rhodamine B likely diffuses close to or at the same rate as the water solvent, in contrast to the Zn-based materials. The normal collisions with water molecules, determined by Brownian motion, are absent and play little role in impeding flow and transport. Generally, the silica surface of the nanoparticles is hydrophilic and at silica/water interfaces negative charge is present on the water side. Such hydrophilic surfaces will be more likely to generate ordering within the solvent[37,38] which goes well beyond the electrical double layer. A further reasonable conclusion would be that Rhodamine B diffusion within such a layer should (at least) have the characteristics associated with ballistic transport[32]. The opposite is also possible – the slow (sub)diffusion of the zinc porphyrin may also have a contribution arising from trapping within potential areas of turbulence away from the ordered layers as percolation of the solvent itself establishes far-from-equilibrium solvent transport conditions.

The subject of long range ordering of water molecules is ill defined in the literature. It can be explained by two mechanisms, either stacked dipole aligned layers or actual structural rearrangement of the water itself, possibly after deprotonation or after surface chemical changes, for example. The role of long range van der Waals forces and dipole alignment, together with charge balance, has been proposed to explain the build-up of hydroxyl groups at air/water interfaces[39] with deprotonation whilst at the silica/water interface local numerical modelling[40] and experimental studies using x-rays[41] support the formation of two types of positively charged defects: transient $SiOH_2^+$ and permanent Si-($OH^+$)-Si at strained Si-O-Si sites. The spectral bands of these sites are consistent with those attributed to ice-like (or graphene like) water under confinement using phase sensitive surface spectroscopy[42]. Together with water protonation to form hydronium ions, $H_3O^+$, away from the negatively charged interface region there is support for the existence of routes to interfacial water deprotonation under confinement and therefore reorientation and ordering of the local structure. Some experiments support the arguments that such ordering can be nanometres thick - up to 8 nm having been measured[43], larger than the pores of the microfibers reported here. With photolytic excitation at appropriate interfaces, long range ordering of hundreds of micron thick, in so-called "exclusion zones", exist[44]. These negatively charged ice-like liquid structures, with a surrounding volume of positively charged hydronium ions (or protons) account for the enhanced lubricating properties of water under confinement[45] analogous to graphene lubricants. Similar liquid water lubricated ice flow within micron-sized silica channels has even been used to infer the presence of laser-induced densification corrugations on the channel surface[46] – it seems probable that the ice-like water structure generated within nanoscale confinement is identical to the super cooled liquid layer pre-existing on ice which can also be a few nanometres thick depending on temperature[47]. Further evidence of charge separation is the observation of enhanced proton conductivity in mesoporous silica[48-50], which has also been exploited in micron sized channels to demonstrate electrokinetic batteries[51] and determine channel potentials[52], procedures that can be used to tune the exclusion zone and therefore potentially affect diffusion rates. Deprotonation of both surface and water is therefore an important part involved in the process of water-assisted percolation so the pH of a solution can also affect structure (including the self-assembly of wires), as can pre-silanation of the silica surface.

This description implies that the addition of a charged, hydrophilic material such as Rhodamine B, which readily releases $Cl^-$ in de-ionised water, can alter the solvent pH, and therefore the degree of ordering compared to the zinc molecules. The negatively charged interfacial water layers therefore repel the anion and attract the Rhodamine, which may also orientate itself via both attraction to the positively charged silica surface and possible interactions between the carboxylic group and silanated species. The introduction of additional charge will also affect the extent of the exclusion zone and may therefore be sensitive to Rhodamine B concentration. In this way, the flow of Rhodamine B is determined primarily by solvent flow with very little impediment from thermal, or Brownian, motion. In statistical terminology, very little, if any, white noise is present. By contrast, the larger two dimensional structure of ZnTPP is largely impeded through aggregation and clogging, as well as potentially being impeded by a reduced cross-sectional flow area as a result of the existence of the exclusion zones. Past work has shown porphyrin-based self-assembled sheets on glass slides have readily formed over each other with no strong attachment to silica[53] suggesting adsorption is not expected to be a major contributor. In the special case of tin (IV) porphyrins, there is evidence that they can attach through chemisorption to the inner silica surface of an air-structured optical fibre microchannel by generation of a Sn-O-Si bridge through charge transfer [54].

The experimental results demonstrate extremely effective filtration of the porphyrin from the water solvent. The potential biological analogue of such water assisted diffusion and filtering are obvious – blood containing iron haemoglobin does not diffuse into cells providing a suitable channel medium to deliver the required species that do diffuse into much of the body. The potential of classical molecular sieves is suggested by the significant differences in percolation between the two Zn compounds used in this work.

## 4. Experimental Section

**Microfibre fabrication:** Long silica microfibres were fabricated by gravity assisted directional evaporative self-assembly ("GADESA") where the substrate upon which the process is undertaken is tilted slightly. The method is largely evaporative self-assembly through convective flow and fracturing upon drying and is similar to that described in reference 3. This generates a preferential convective flow in one axis within the larger drops, which are flat-topped rather than spherical, helping to generate more uniform and aligned microfibres during evaporation and subsequent fracturing. A solution of ~ 5 wt% of silica nanoparticles in water, diluted from a 40 wt% colloidal suspension (pH = 9.1) obtained from Sigma Aldrich, was deposited by pipette onto a glass substrate with a small tilt away from the horizontal position so that the drop slowly moved towards one end spreading itself out asymmetrically. The size distribution of the nanoparticles has, in previous work[1], been measured by dynamic light scattering (DLS) to be predominantly ($\phi \sim 20 - 30$) nm. GADESA breaks the otherwise uniform radial microfluidic flow of particles towards the drop edge and, in addition to preferential flow, leads to preferential stresses along one axis generating longer wires from end to end. In contrast to previous work using lasers to generate non-uniform drops[2] to create uniform wires within small drops, this method is better suited to the fabrication of large quantities of longer wires. Upon drying uniform wires (taper aspect ratio TAR = 1) up to record lengths of 11.5 cm and cross-sections of $w \times h$ = 40 μm x 10 μm were formed. The similar charge between the microfibre and surface leads to ready detachment of the microfibres.

**Optical diffusion measurement:** For these experiments, the red ($\lambda = 632.5$ nm) output from a HeNe laser is launched into a standard telecoms optical fibre which is then coupled at a slight offset into the microfibre (Figure 4) – the light then spreads to fill out the microfibre. This form of single-mode to multimode coupling risks turning the slab waveguide into a multimode interference device (MMI)[24]. This makes a diffusion analysis difficult. The large number of modes present in the highly multimode microfibre will be equally excited by cross-coupling and, assuming no changes in refractive index, average out interference effects over the length of region into which a sample percolates. The output light is collected using a microscope objective and collimated towards a fast photodetector/power meter.

In the first experiment, a drop of Rhodamine B in water ([Rh B] ~ 0.00135M) was placed on the surface of the wire from one side and transmission through the wire was monitored on the power meter. Its size lies within the two sites of hcp structure; faster diffusion is expected to occur through the larger pore sizes. Although capillary action spreads the drop around the wire, the large rectangular dimensions (~ 40 μm width, 10 μm height) can, within a first approximation, allow the problem to be treated as predominantly one sided diffusion from the top flat face. More accurately, penetration from all sides will lead towards an approximately quadratic profile (for an ideal circular cross-section) convolved with the diffusion profile.

When the drop is placed onto the wire a small perturbation from coupling mismatch is expected given how thin these microfibres are. This will affect input coupling from the single mode telecoms fibre. For reference, deionized water was also placed on the wires in an identical manner several times to rule out attenuation induced by other factors and to ensure that the process leads to minimal perturbation.

**Laser ablation inductive coupled mass spectrometry (LA-ICP-MS):** Once the two samples are prepared, the wires are placed in the instrument chamber. A frequency quintupled, Q-switched Nd:YAG laser is focussed and scanned across the sample surface ($\lambda = 213$ nm, $f_{pulse} = 60$ kJ/cm$^2$, repetition rate = 10 Hz, $w_0 = 8$ μm, $v = 5$ μm/s). Material is ablated at this point and transported by a carrier gas to a mass spectrometer for analysis of content. The effective threshold is $E_{ab} < 398$ mJ/cm$^2$ at 213 nm. Ar was used as the carrier gas to maintain plasma stability. This ablation was found not to be entirely uniform, and occurred preferentially at the edges, which adds some uncertainty to the derivations obtained with data analysis. It is explained in part by remaining stresses and any edge non-uniformities that generate hot spots within the wire. The laser is scanned back and forth to accumulate ablation in steps; within error the ablation as a function of passes is close to linear. For these experiments, the scanning rate is 5 μm/s, so it takes $t \sim (8-9)$s to traverse the waveguide in a single pass.

## 5. Conclusions

In conclusion, the percolation diffusion of three molecules into optically transparent self-assembled silica microfibers has been studied. At opposite ends, optical transmission experiments showed rapid "super" diffusion of the Rhodamine B but no detectable diffusion of ZnTPP. The signal oscillations observed with Rhodamine B are attributed to complex waveguide mode behaviour as the local index rises with diffusing water (from $n \sim 1.35$ to 1.42). The absence of penetration of the ZnTPP was confirmed using laser ablation inductive coupled mass spectrometry (LA-ICP-MS) where the diffusion is found to be extremely slow (but non-zero). Zinc acetate dehydrate diffuses faster than the porphyrin, but much slower than the Rhodamine B, despite comparable sizes, overall demonstrating the potential of molecular sieves using these microfibres. The exact nature for this selectivity is likely to involve both charge and dipole and structure assisted enhancement of diffusion, an interesting feature given the likely role of dipole ordering of water in biological diffusive and osmotic cell processes, for example. The diffusion of the Zn(II) species appears as a simple exponential consistent with slow molecular diffusion much slower than solvent penetration. Although diffusion generally appears to be close to exponential, the timescales involved for Rhodamine B point to it penetrating the system as fast as the solvent, de-ionized water – in this case the transport system is far from equilibrium and the normal diffusive analysis may not apply. This rapid penetration stands out against the similarly sized zinc compounds. Broadly, the results of this work point to marked differences in diffusion into mesoporous structures based on electrostatic and intermolecular forces of different species and this in of itself can aid the design of micro and nanofluidic devices and sensors. Given the principles explored here apply to both two and three dimensional self-assembly, this includes biomedical applications such as filtering and separation, protein trapping and preservation[55] as well as drug release capsules using mesoporous silica[56,57].

## Acknowledgments

The authors acknowledge support from the Australian Research Council (ARC) through grants ARC FT110100116. M. Ma acknowledges an *i*PL summer scholarship.

## Conflicts of Interest

The authors declare no conflict of interest.